\documentclass[12pt]{article}
\usepackage{amssymb,amsmath}
\usepackage[noblocks]{authblk}
\usepackage[top=0.75in, bottom=0.75in, left=0.75in, right=0.75in, dvips]{geometry}
\usepackage{caption}
\usepackage{graphicx}
\usepackage{epstopdf}
\date{}
\begin{document}
\textwidth 10.0in
\textheight 9.0in
\topmargin -0.60in
\title{Multiple Couplings and Renormalization Scheme Ambiguities}
\author[1,2]{D.G.C. McKeon}
\author[1]{Chenguang Zhao}
\affil[1] {Department of Applied Mathematics, The
University of Western Ontario, London, ON N6A 5B7, Canada}
\affil[2] {Department of Mathematics and
Computer Science, Algoma University,\newline Sault Ste. Marie, ON P6A
2G4, Canada}
  
\maketitle                                 
   
\maketitle
\noindent
email: dgmckeo2@uwo.ca

\begin{abstract}
The ambiguities inherent in renormalization are considered when using mass independent renormalization in massless theories that involve two coupling constants.  We review how unlike models in which there is just one coupling constant there is no renormalization scheme in which the $\beta$-functions can be chosen to vanish beyond a certain order in perturbation theory, and also the $\beta$-functions always contain ambiguities beyond first order. We examine how the coupling constants depend on the coefficients of the $\beta$-functions beyond one loop order. A way of characterizing renormalization schemes that doesn't use coefficients of the $\beta$-function is considered for models with either one or two couplings.  The renormalization scheme ambiguities of physical quantities computed to finite order in perturbation theory are also examined. The renormalization group equation makes it possible to sum the logarithms that have explicit dependence on the renormalization scale parameter $\mu$ in a physical quantity R and this leads to a cancellation with the implicit dependence of R on $\mu$ through the running couplings, thereby removing the ambiguity associated with the renormalization scale parameter $\mu$. It is also shown that there exists a renormalization scheme in which all radiative contributions beyond lowest order to R are incorporated into the behavior of the running couplings and the perturbative expansion for R is a finite series.
\end{abstract}

\section{Introduction}

In order to excise divergences arising in the perturbative evaluation of physical quantities using quantum field theory, it is necessary to perform a subtraction to ``renormalize'' the parameters that characterize the theory\footnote{Analytic continuation can be used to avoid explicit occurrence of divergences [1].}. Ambiguities in perturbative results arise both from the introduction of an unphysical scale parameter $\mu$ and from the possibility of performing a finite renormalization in addition to what is required to eliminate the divergence.  The requirement that the exact expression for physical quantities be unambiguous leads to the renormalization group (RG) equations [2-4].

The renormalization scheme (RS) ambiguities when one uses a mass independent RS [5,6] in theories with a single coupling constant $a$ can be parameterized by the coefficients $c_i(i \geq 2)$ of the RG $\beta$-function that arise beyond two loop order, with the one and two loop coefficients being RS invariant [7].  It is possible to find a function $B_i(a, c_k)$ that shows how this coupling $a$ depends on these coefficients $c_i$ [8].  Furthermore, it is possible to use the RG equation associated with $\mu$ to sum these terms which in perturbation theory explicitly depend on $\mu$ through $\ln \mu$ so that this explicit dependence of a physical quantity $R$ on $\mu$ cancels against implicit dependence on $\mu$ through $a(\mu)$ [9-11].  

In this paper, we extend these considerations to deal with the situation in which there are two coupling constants in a massless theory. It turns out there are significant differences when one goes from one to two couplings. We first review how when using mass independent renormalization the $\beta$-functions associated with these couplings are RS dependent at two loop order and beyond.  This is unlike the situation in which there is only one coupling where at two loop order the $\beta$-function is RS independent. (This has been noted in ref. [12] and again in ref. [19].)

A second feature of a theory in which there are two couplings is that, unlike the situation in which there is but one coupling, there is no RS in which the $\beta$-functions can be terminated beyond two loop order.  When there is only one coupling, the $\beta$-function receives only one and two loop contributions when the 't Hooft [13] RS is used.

At N loop order the $\beta$-functions in a model with two couplings involve $2(N+2)$ parameters.  We show how the RS used can be characterized by $2(N+1)$ of these parameters; in general the two other parameters are dependent on these $2(N+1)$ parameters.  This motivates developing a way of characterizing a RS by use of parameters that arise in the expansion of the coupling in one RS in terms of the coupling in another RS.

The RS dependence of perturbative expressions for a physical quantity $R$ is considered when there are two couplings.  It is demonstrated how $R$ is independent of $\mu$ when RG summation is performed and once this is done, how $R$ depends on parameters that characterize the change in RS.

It is shown that when there are either one or two coupling constants, a RS can be chosen so that the perturbative expansion for $R$ terminates and the effect of all higher loop effects is absorbed into the behavior of the running coupling.

In the next section we review some features of RS dependence when there is one coupling.  By way of contrast, the analogous results when there are two couplings are presented.

In this paper we do not consider models when there are more than two couplings. However, we believe that there is no significant qualitative difference in the RS dependence of models with two and models with more than two couplings. As noted above though, there are qualitative differences in the RS dependence of models with one and models with two couplings.

We wish to emphasize that we are exclusively using mass independent renormalization schemes. When using a mass dependent RS, there are non-trivial differences in the RS ambiguities in the theory [26,27].

\section{Renormalization Scheme Dependence With One Coupling}

Quantum chromodynamics (QCD) is characterized by a single couplant $a$.  When using the notation of ref. [8], the dependency of $a$ on the renormalization scale parameter $\mu$ is given by
\begin{align}
\mu \frac{da}{d\mu} &= \beta(a)\nonumber \\
&= - ba^2 \left( 1 + ca + c_2 a^2 + \ldots \right)
\end{align}
when using a mass independent renormalization scheme [5,6].

If a finite renormalization is performed [12], then
\begin{equation}
\overline{a} = a + x_2 a^2 + x_3 a^3 + \ldots
\end{equation}
obeys an equation like (1).  We find that since
\begin{subequations}
\begin{align}
\mu \frac{d\overline{a}}{d\mu} &= \beta(a) \left(1 + 2x_2 a + 3x_3 a^2 + \ldots\right)   
\intertext{ as well as }
& = - \overline{b}\left(a + x_2 a^2 + x_3 a^3 + \ldots\right)^2 \Big[ 1 + \overline{c} \left(a + x_2 a^2 + \ldots \right)  \\
&\hspace{2cm} + \overline{c}_2 \left(a + x_2a^2 + \ldots\right)^2 + \ldots \Big]\nonumber
\end{align}
\end{subequations}
then by eqs. (3a, 3b) we find that [23]
\begin{subequations}
\begin{align}
\overline{b} &= b\\
\overline{c} &= c\\
\overline{c}_2 = c_2 - & cx_2 + x_3 - x_2^2\\
\overline{c}_3 = c_3 - 3cx_2^2 + & 2\left(c_2 - 2\overline{c}_2\right) x_2 + 2x_4 - 2x_2 x_3\\
\overline{c}_4 = c_4 - 2 x_4x_2 - x_2^3 + & c\left(x_4 - x_2^3 - 6x_2x_3\right) + 3x_3c_2 - 4x_3\overline{c}_2\\
&- 6x_2^2 \overline{c}_2 + 2x_2c_3 - 5x_2\overline{c}_3 + 3x_5\nonumber
\end{align}
\end{subequations}
etc.

From eqs. (4a-4e) we find
\begin{subequations}
\begin{align}
& x_3 = \overline{c}_2 - c_2 + cx_2 + x_2^2\\
& x_4 = \frac{1}{2}\left[ \overline{c}_3 - c_3 + \left(6 \overline{c}_2 -
4c_2\right) x_2 + 5c x_2^2 + 2x_2^3\right]\\
& x_5 = \frac{1}{3}\Big\{ \overline{c}_4 - c_4 + \left(5x_2 \overline{c}_3 -
2x_2 c_3\right) + \left(4\overline{c}_2 - 3c_2 + 6x_2 c\right)\\
&\hspace{1cm}\left( \overline{c}_2 - c_2 + c x_2 + x_2^2\right) + \left( \overline{c}_2 - c_2 + 
c x_2 + x_2^2\right)^2 + 6x_2^2 \overline{c}_2\nonumber \\
&\hspace{1.2cm}+ x_2^3 c + \left(2x_2 -c \right) \Big[ \frac{1}{2} \left(\overline{c}_3 - c_3 \right) + x_2 \left(3\overline{c}_2 -2 c_2 \right)\nonumber \\
&\hspace{1.4cm}+ \frac{5}{2} cx_2^2 + x_2^3 \Big] \Big\} \nonumber
\end{align}
\end{subequations}
etc.\\
We see that the renormalization of $a$ in eq. (2) leads to a change in $c_i(i \geq 2)$ that fix $x_i(i \geq 3)$ with $x_2$ not determined. In ref. [23,24], some restrictions on the transformation of eq. (2) are considered.

The fact that a RS is characterized by $c_i$ means that $a$ itself is dependent on $c_i$.  If
\begin{equation}
\frac{da}{dc_i} = B_i(a, c_k)
\end{equation}
then the function $B_i$ can be determined by the consistency condition
\begin{equation}
\left[ \mu \frac{\partial}{\partial\mu}, \frac{\partial}{\partial c_i}\right] a = 0
\end{equation}
which leads to [8]
\begin{align}
B_i(a,c_k) &= -b\beta(a) \int_0^a dx \frac{x^{i+2}}{\beta^2(x)}\nonumber \\
& \approx \frac{a^{i+1}}{i-1} \left[ 1 + \left( \frac{(-i+2)c}{i}\right) a + \left( \frac{(i^2-3i+2)c^2(-i^2+3i)c_2}{(i+1)i}\right)a^2 + \ldots\right].
\end{align}
If now
\begin{equation}
\mu \frac{d}{d\mu}a \left(\overline{\mu}, c_i\right) = 0 = \left(
\mu \frac{\partial}{\partial\mu} + \beta(a) \frac{\partial}{\partial a}\right)
\left( a \left(\mu , c_i\right) + \left(\sigma_{21} \ell \right) a^2 
\left(\mu , c_i\right) + \ldots \right) \quad \left( \ell \equiv b \ln \left(\frac{\mu}{\overline{\mu}}\right)\right)
\end{equation}
then we have
\begin{align}
& \sigma_{21} = 1, \quad \sigma_{31} = c, \quad \sigma_{32} = 1, \quad \sigma_{41} = c_2, \quad \sigma_{42} = \frac{5}{2}c, \quad \sigma_{43} = 1\\
& \sigma_{51} = c_3, \quad \sigma_{52} = 3c_2 + \frac{3}{2} c^2,  \quad \sigma_{53} = \frac{13}{3}c, \quad \sigma_{54} = 1\nonumber \\
& \sigma_{61} = c_4, \quad \sigma_{62} = \frac{7}{2}\left(cc_2 + c_3\right), \quad
\sigma_{63} = \frac{1}{6} \left(35c^2 + 36c_2\right), \quad \sigma_{64} = \frac{77}{12} c, \quad \sigma_{65} = 1\nonumber \\
& \sigma_{71} = c_5, \quad \sigma_{72} = 2\left(c_2^2 + 2cc_3 + 2c_4\right), \quad
\sigma_{73} = \frac{1}{6} \left(15c^2 + 92cc_2 + 48 c_3\right), \nonumber \\ & \hspace{2.4cm}\sigma_{74} = \frac{5}{6} \left( 17c^2 + 12 c_2\right), \quad \sigma_{75} = \frac{87}{10} c, \quad \sigma_{76} = 1.\nonumber 
\end{align}
Knowing these coefficients $\sigma_{mn}$ gives $a(\bar{\mu},c_i)$ in terms of $a(\mu,c_i)$; this amounts to having a perturbative solution of eq. (1) [28]. If one defines $S_n (a)=\sum_{k=0}^{\infty}\sigma_{k+n+1,k}a^{k+n+1}$ (n=0,1,2...), one can solve sequentially for $S_n$ using eq. (9).

Similarly, if
\begin{equation}
a \left(\mu, \overline{c}_k\right) = a\left(\mu , c_k \right) + \lambda_2 \left( c_k, \overline{c}_k\right) a^2 \left(\mu, c_k\right) + \lambda_3 \left( c_k, \overline{c}_k\right) a^3 \left(\mu, c_k\right) + \ldots
\end{equation}
with $\lambda_i (c_k, c_k) = 0$, then the equation
\begin{equation}
\frac{d}{dc_i} a \left(\mu, \overline{c}_k\right) = 0 = 
\left( \frac{\partial}{\partial c_i} + B_i \left( a, c_k\right) \frac{\partial}{\partial a}\right) \left( a(\mu , c_k) + \lambda_2\left( c_k, \overline{c}_k\right) a^2 \left(\mu, c_k \right) + \ldots \right)
\end{equation}
results in
\begin{equation}
\lambda_2 = (\overline{c}_2 - c_2),\quad   \lambda_3 = \frac{1}{2}(\overline{c}_3 - c_3),\quad \lambda_4 = \frac{1}{6}\left(\overline{c}_2^2 - c_2^2\right) + \frac{3}{2}
(\overline{c}_2 - c_2) - \frac{c}{6}\left(\overline{c}_3 - c_3\right) + \frac{1}{3}
(\overline{c}_4 - c_4)
\end{equation}
etc.  

Eqs. (11-13) is essentially a series solution of eq. (6) [28].

If in eq. (2) we eliminate $x_n(n \geq 3)$ in favour of $x_2$, $c_i$, $\overline{c}_i (i \geq 2)$ using eq. (5) and then set $c_i = \overline{c}_i$, we end up with the series of eq. (9) for $a(\overline{\mu}, c_i)$ provided $x_2 = b \ln \left( \frac{\mu}{\overline{\mu}}\right)$ [11]. This shows that $x_2$ can be identified with $b\ln\left(\frac{\mu}{\overline{\mu}}\right)$ as postulated in ref. [8].

If now a physical quantity, such as the cross section $R_{e^+e^-}$ for $e^+e^- \longrightarrow$ (hadrons), is expanded in the form
\begin{equation}
R = \sum_{n=0}^\infty A_n(a) L^n
\end{equation}
where $L = b\ln \frac{\mu}{Q}$ and [9, 10]
\begin{equation}
A_n(a) = \sum_{k=0}^\infty T_{n+k,n} a^{n+k+1},
\end{equation}
then from the RG equation
\begin{equation}
\left( \mu \frac{\partial}{\partial \mu} + \beta(a) \frac{\partial}{\partial a}\right) R = 0\nonumber
\end{equation}
it follows that
\begin{equation}
A_n(a) = - \frac{\beta(a)}{bn} \frac{d}{da}	 A_{n-1}(a)
\end{equation}
so that since by eq. (1) [8]
\begin{equation}
\ln \left( \frac{\mu}{\Lambda}\right) = \int_0^{a \left(\ln \frac{\mu}{\Lambda}\right)} \frac{dx}{\beta(x)} + \int_0^\infty \frac{dx}{bx^2(1+cx)}
\end{equation}
we find from eqs. (14-17)
\begin{equation}
R = A_0 \left( a\left( \ln \frac{Q}{\Lambda}\right)\right)
\end{equation}
and the explicit and implicit dependence of $R$ on the unphysical scale parameter $\mu$ has cancelled [10].

By eqs. (15, 18) we see that
\begin{equation}
R = \sum_{n=0}^\infty T_n\left( a \left(\ln \frac{Q}{\Lambda}\right)\right)^{n+1} \quad \left( T_n \equiv T_{n,0}\right).
\end{equation}
so that from the requirement that
\begin{equation}
\left( \frac{\partial}{\partial c_i} + B_i(a) \frac{\partial}{\partial a}\right) R = 0
\end{equation}
we find that 
\begin{align}
& T_0 = \tau_0, \quad T_1 = \tau_1, \quad T_2 = -c_2 + \tau_2, \quad T_3 = -2c_2 \tau_1 - \frac{1}{2} c_3 + \tau_3\\
&\hspace{2cm} T_4 = -\frac{1}{3}  c_4 - \frac{c_3}{2}\left( - \frac{1}{3} c + 2\tau_i\right) + \frac{4}{3} c_2^2 - 3 c_2\tau_2 + \tau_4 \nonumber
\end{align}
etc.\\
where the $\tau_i$ are constants of integration and hence are RS invariants [9, 10].  One RS of particular interest is the one in which $T_i = 0 (i \geq 2)$ so that R is represented by a perturbative series that terminates. A second interesting RS due to 't Hooft has $c_i = 0 ( i \geq 2)$ [13, 14], so that the $\beta$ function is a finite series in the coupling.

We will now see how the results obtained in this section are modified when one considers models in which there are two coupling constants. Again, we will deal with massless theories and employ mass independent renormalization schemes.

\section{ Renormalization Scheme Dependence With Two Couplings}

We now will consider the consequences of having two couplings $g_a(a = 1,2)$ in a model with the $\beta$-functions
\begin{equation}
\mu \frac{dg_a}{d\mu} = \beta_a (g_1,g_2) = \sum_{i=2}^\infty \sum_{j=0}^i c_{ij}^a(g_1)^{i-j} (g_2)^j
\end{equation}
in place of eq. (1). In order to compute $c^a_{ij}$, a calculation of diagrams involving i-1 loops is required. For example, in the limit of the Standard Model in which there is only the $SU(2)$ gauge field and the Higgs doublet, with the gauge coupling $g$ and the Higgs self coupling $\lambda$, the coefficient $c^a_{ij}$ are to two loop order [15] in the $\overline{MS}$ RS if $g^2 = 16\pi^2g_1$ and $\lambda = 16\pi^2g_2$,
\begin{align}
c_{20}^1 &= -\frac{19}{3}, \quad c_{30}^1 = \frac{35}{3}, \quad c_{20}^2 = \frac{27}{4}, \quad c_{21}^2 = -9, \quad c_{22}^2 = 4\\
c_{30}^2 &= \frac{915}{8}, \quad c_{31}^2 = -\frac{73}{8}, \quad c_{32}^2 = 18, \quad c_{33}^2 = -\frac{26}{3}\nonumber .
\end{align}

The analogue to eq. (2) for a finite renormalization of $g_a$ is
\begin{equation}
\overline{g}_a = g_a + \sum_{i=2}^\infty \sum_{j=0}^i x_{ij}^a (g_1)^{i-j} (g_2)^j.
\end{equation}
In analogy with eqs. (3a) and (3b) we then see that
\begin{subequations}
\begin{align}
\mu \frac{d\overline{g}_a}{d\mu} &= \beta_a (g_b) + \sum_{i=2}^\infty \sum_{j=0}^i x_{ij}^a  \left[ (i-j) g_1^{i-j-1} g_2^j \beta_1 (g_b)
 + jg_1^{i-j} g_2^{j-1} \beta_2 (g_b)\right]\\
 \intertext{and} 
 \mu \frac{d\overline{g}_a}{d\mu} &= \sum_{i=2}^\infty \sum_{j=0}^i \overline{c}_{ij}^a  \left[ g_1 + \sum_{k=2}^\infty \sum_{\ell=0}^k x^1_{k\ell} g_1^{k-\ell} g_2^\ell\right]^{i-j} \left[ g_2 + \sum_{m=2}^\infty \sum_{n=0}^m x^2_{mn}  g_1^{m-n} g_2^n\right]^j.
\end{align}
\end{subequations}
Upon comparing terms in eqs. (25a) and (25b) that are quadratic in the couplings (ie, that are of one loop order) we find that much like eq. (4a)
\begin{equation}
\overline{c}_{2j}^a = c_{2j}^a \quad (j = 0, 1, 2)
\end{equation}
and so one loop contributions to $\beta_a(g_b)$ are RS independent.  However, terms in eqs. (25a) and (25b) that are cubic in the couplings (ie, that are at two loop order) show that at order $g_2^3$, $g_2^3$, $g_1^2g_2$ and $g_1g_2^2$ respectively [19]
\begin{subequations}
\begin{align}
\overline{c}_{30}^a &= c_{30}^a + 2x_{20}^a c_{20}^1 - 2c_{20}^a x_{20}^1 + x_{21}^a c_{20}^2 - c_{21}^a x_{20}^2 \\
\overline{c}_{33}^a &= c_{33}^a + 2x_{22}^a c_{22}^2 - 2c_{22}^a x_{22}^2 + x_{21}^a c_{22}^1 - c_{21}^a x_{22}^1 \\
\overline{c}_{31}^a &= c_{31}^a + 2x_{20}^a c_{21}^1 - 2c_{20}^a x_{21}^1 + x_{21}^a 
\left(c_{20}^1 + c_{21}^2\right) - c_{21}^a \left(x_{20}^1 + x_{21}^2\right)
+ 2x^a_{22} c_{20}^2 - 2c_{22}^a x_{20}^2 \\
\overline{c}_{32}^a &= c_{32}^a + 2x_{20}^a c_{22}^1 - 2c_{20}^a x_{22}^1 + x_{21}^a 
\left(c_{21}^1 + c_{22}^2\right) - c_{21}^a \left(x_{21}^1 + x_{22}^2\right)
+ 2x^a_{22} c_{21}^2 - 2c_{22}^a x_{21}^2 
\end{align}
\end{subequations}
with $a = 1,2$.  From eq. (27) it is immediately apparent that the two loop contributions to $\beta_a(g_1,g_2)$ are RS dependent, unlike what happens when there is one coupling (see. eq. (4b)) [12, 19].  However, as there are now eight equations fixing changes in the eight quantities $c_{3i}^a (a = 1,2; i = 0,1,2,3)$ in terms of just the six independent coefficients $x_{2i}^a (a = 1,2; i = 0,1,2)$, it is evident that it is in general not possible to vary each of the quantities $c_{3i}^a$ independently.  Only if the coefficients $c_{2i}^a (a = 1,2; i = 0,1,2)$ were to have special values would it be possible to find values of $x_{2i}^a$ so that each of the $c_{3i}^a$ equals zero, which would be the analogue of the 't Hooft RS when there is one coupling [13,14].

When one goes beyond two loop order, equations much like eq. (27) can be found. At $N$ loop order, $\overline{c}^a_{N+1,i} - c^a_{N+1,i} (a = 1,2; i = 0 \ldots N + 1)$ is related to $x_{N,i}^a (a = 1,2; i = 0 \ldots N)$ through $2(N+2)$ equations.  Consequently, in general, 2 of the $2(N+2)$ quantities $c_{N+1,i}^a$ cannot be varied independently by altering the RS by adjusting only the $2(N+1)$ independent parameters $x_{N,i}^a$.  However, there is the intriguing possibility that for some choice of 
$x_{i,j}^a$ that either $\beta_1(g_1,g_2)$ or $\beta_2(g_1,g_2)$ vanishes beyond one loop order.

Since not all of the coefficients $c_{mn}^a$ can be varied independently by a change of $RS$, it is apparent that these coefficients are no longer suitable for characterizing a RS where there is more than one coupling.  In the next section we show how the coefficients $x_i$ in eq. (2) (when there is one coupling) or $x_{ij}^a$ in eq. (24) (when there are two couplings), all of which are independent, can be used to characterize a RS.

RS ambiguities are of practical importance, as is illustrated by the discrepancy between the calculations presented in refs. [20] and [21].  This is discussed in ref. [22].

\section{An Alternate Way to Characterize a Renormalization Scheme}

We begin by considering the case of one coupling $a$ and showing how the parameters $x_i$ in eq. (2) can be used to characterize a RS.  Suppose that $a$ refers to the coupling in some ``base scheme'' such as $\overline{MS}$, and the $\overline{a}$ is the coupling in some other scheme with $a$ and $\overline{a}$ related by eq. (2).  If now
\begin{equation}
a = \overline{a} + y_2 \overline{a}^2 + y_3 \overline{a}^3 + \ldots
\end{equation}
then eqs. (2, 28) are consistent provided
\begin{equation}
a = \overline{a} - x_2 \overline{a}^2 + \left( 2x_2^2 - x_3\right)\overline{a}^3 + \left(5x_2 x_3 - 5x_2^3 - x_4\right) \overline{a}^4 + \ldots .
\end{equation}

It is clear that $\overline{a}$ depends on $x_i$; from eq. (2) we see that
\begin{subequations}
\begin{align}
\frac{d\overline{a}}{dx_n} &= a^n \qquad (\overline{a}(x_n = 0) = a)\\
\intertext{ which by eq. (29) becomes}
\frac{d\overline{a}}{dx_n} &\equiv \overline{B}_n (\overline{a}, x_m) = 
\left( \overline{a}- x_2\overline{a}^2 + \left( 2x_2^2 - x_3\right) \overline{a}^3 + \ldots \right)^n.
\end{align}
\end{subequations}
There are two consistency checks on eq. (30b).  First of all, we have
\begin{subequations}
\begin{align}
&\hspace{5cm}\frac{da}{dx_n} = 0\\
\intertext{which by (29) and (30b) leads to}
&\left[ \frac{\partial}{\partial x_n} + \left( \overline{a} - x_2 \overline{a}^2 + \left( 2x_2^2 - x_3\right)\overline{a}^3 + \ldots\right)^n \frac{\partial}{\partial\overline{a}}\right] \left( \overline{a} - x_2 \overline{a}^2 + \left( 2x_2^2 - x_3 \right) \overline{a}^3 + \ldots \right) = 0
\end{align}
\end{subequations}
which can be verified.  A second test follows from eq. (1)
\begin{subequations}
\begin{align}
\mu \frac{da}{d\mu} &= -ba^2 ( 1 + ca + c_2a^2 + \ldots );\\
\intertext{if we eliminate $a$ in eq. (32a) by eq. (29) and use}
\mu \frac{d\overline{a}}{d\mu} &= -\overline{b}\overline{a}^2 ( 1 + \overline{c}\;\,
\overline{a} + \overline{c}_2 \overline{a}^2 + \ldots )
\end{align}
\end{subequations}
we recover eq. (4).

We can now employ this approach to characterizing a RS to the situation in which there are two couplings.  In this case, a RS is defined in terms of a ``base scheme'' in which the couplings are given by $(g_1, g_2)$ and the coefficients $x_{mn}^a$ appearing in eq. (24).  The advantage of this approach is that all of the $x_{mn}^a$ can be independently varied.  We have shown that it is not possible to independently vary the coefficients $c_{ij}^a$ appearing in the functions $\beta_a$ in eq. (22) by use of eq. (2).

We begin by noting that from eq. (24), it follows that if
\begin{equation}
g_a = \overline{g}_a + \sum_{m=2}^\infty \sum_{n=0}^m Y_{mn}^a \overline{g}_1^{\;m-n} \;\overline{g}_2^n
\end{equation}
then
\begin{equation}
Y_{2k}^a + x_{2k}^a = 0 \quad (a = 1,2; k = 0,1,2)
\end{equation}
and
\begin{equation}\tag{35a,b}
 Y_{30}^1 = 2(x_{20}^1)^2 + x_{21}^1 x_{20}^2 - x_{30}^1; \quad
 Y_{33}^2 = 2(x_{22}^2)^2 + 2x_{21}^2 x_{22}^1 - x_{33}^2
\end{equation}
\begin{equation}\tag{35c,d}
 Y_{33}^1 = 2x_{22}^1 x_{22}^2 + x_{21}^1 x_{22}^1 - x_{33}^1; \quad
 Y_{30}^2 = 2x_{20}^2 x_{20}^1 + x_{21}^2 x_{20}^2 - x_{30}^2 
\end{equation}
\begin{equation}\tag{35e,f}
 Y_{31}^1 = 2x_{20}^1 x_{21}^1 + x_{21}^1 (x_{20}^1 + x_{21}^2) + 2x_{22}^1 x_{20}^2 - x_{31}^1; \;\;
 Y_{32}^2 = 2x_{22}^2 x_{21}^2 + x_{21}^2 (x_{22}^2 + x_{21}^1) + 2x_{20}^2 x_{22}^1 - x_{32}^2
\end{equation}
\begin{equation}\tag{35g,h}
Y_{32}^1 = 2x_{22}^1 x_{21}^2 + x_{21}^1 (x_{22}^2 + x_{21}^1) + 2x_{20}^1 x_{22}^1 - x_{32}^1;  \;\;
 Y_{31}^2 = 2x^2_{20} x_{21}^1 + x_{21}^2 (x_{20}^1 + x_{21}^2) + 2x_{22}^2 x_{20}^2 - x_{31}^2.
\end{equation}
etc.\\
The inversion of series with several variables is discussed in, for example, ref. [18].

It also follows from eq. (24) that
\begin{equation}\tag{36}
\frac{d\overline{g}_a}{dx_{mn}^b} \equiv \overline{B}_{b;m,n}^a (\overline{g}_a) = \delta_b^a g_1^{m-n}g_2^n
\end{equation}
so that, for example
\begin{align}\tag{37}
\frac{d\overline{g}_1}{dx_{21}^1} = g_1g_2 = \overline{g}_1\overline{g}_2 &- 
x_{20}^2 \overline{g}_1^3 - \left( x_{20}^1 + x_{21}^2 \right)\overline{g}_1^2\overline{g}_2 \\
& - \left( x_{21}^1 + x_{22}^2\right) \overline{g}_1\overline{g}_2^2 - x_{22}^1 \overline{g}_2^3 \ldots\nonumber
\end{align}

We now can consider the RS dependence of a physical quantity using the parameters $x_n$ when there is one coupling $a$ and $x_{mn}^a$ when there are two couplings $g_1$, $g_2$.

Again considering $R$ given by eq. (19), we take $a\left(\ln \frac{Q}{\Lambda}\right)$ to be the coupling in a ``base scheme'' (such as $\overline{MS}$).  Under a renormalization such as in eq. (2) we must have
\begin{equation}\tag{38}
\frac{d}{dx_n}R = 0 = \left( \frac{\partial}{\partial x_n} + \overline{B}_n (\overline{a}, x_m)\frac{\partial}{\partial \overline{a}} \right) \sum_{n=0}^\infty \overline{T}_n \left( \overline{a}\left(\ln \frac{Q}{\Lambda}\right)\right)^{n+1}.
\end{equation}
In eq. (38), $\overline{T}_n \equiv \overline{T}_{n,0}$ are the coefficients of an expansion of $R$ in powers of $\overline{a}$, a coupling related to the coupling $a$ through the renormalization of eq. (2).  Using eq. (30b), we find that for $k= 2,3 \ldots$
\begin{equation}\tag{39}
\sum_{n=0}^\infty \left[ \frac{\partial \overline{T}_n}{\partial x_k} \overline{a}^{n+1} + (n+1)\overline{a}^n \left(\overline{a} - x_2\overline{a}^2 + \left(2x_2^2 - x_3\right)\overline{a}^3 + \ldots\right)^k \overline{T}_n\right] = 0.
\end{equation}
From eq. (39) it follows that
\begin{equation}\tag{40a-c}
\frac{\partial \overline{T}_0}{\partial x_2} = \frac{\partial \overline{T}_0}{\partial x_3} = \frac{\partial \overline{T}_0}{\partial x_4} = 0
\end{equation}
\begin{equation}\tag{41a-c}
\frac{\partial \overline{T}_1}{\partial x_2} + \overline{T}_0 = \frac{\partial \overline{T}_1}{\partial x_3} = \frac{\partial \overline{T}_1}{\partial x_4} = 0
\end{equation}
\begin{equation}\tag{42a-c}
\frac{\partial \overline{T}_2}{\partial x_2} + 2 \overline{T}_1 - 2x_2 \overline{T}_0 = \frac{\partial \overline{T}_2}{\partial x_3} + \overline{T}_0 = \frac{\partial \overline{T}_2}{\partial x_4} = 0
\end{equation}
\begin{equation}\tag{43a-c}
\frac{\partial \overline{T}_3}{\partial x_2} + 3 \overline{T}_2 - 4x_2 \overline{T}_1 + \overline{T}_0\left(5x_2^2 - 2x_3\right) = \frac{\partial \overline{T}_3}{\partial x_3} + 2\overline{T}_1 - 3x_2 \overline{T}_0 = \frac{\partial \overline{T}_3}{\partial x_4} + \overline{T}_0 = 0. 
\end{equation}
Since when $x_i = 0$, $\overline{a} = a$ and $\overline{T}_n = T_n$, we see that from eqs. (40-43) that
\begin{equation}\tag{44a}
\overline{T}_0 = T_0
\end{equation}
\begin{equation}\tag{44b}
\overline{T}_1 = T_1 - x_2 T_0
\end{equation}
\begin{equation}\tag{44c}
\overline{T}_2 = T_2  + \left( -x_3 + 2x_2^2\right) T_0 + (-2x_2)T_1
\end{equation}
\begin{equation}\tag{44d}
\overline{T}_3 = T_3  + \left( -x_4 + 5x_2x_3 - 5x_2^3\right) T_0 + \left(-2x_3 + 5x_2^2\right)T_1 - 3x_2T_2
\end{equation}
etc.\\
One interesting feature of eq. (44) is that $x_2, x_3 \ldots$ can all be selected so that $\overline{T}_1 = \overline{T}_2 = \overline{T}_3 \dots = 0$, leaving $R$ given by the single term
\begin{equation}\tag{45}
R = T_0 \overline{a} \left( \ln \frac{Q}{\Lambda}\right).
\end{equation}
In eq. (45), $\overline{a}$ runs according to eq. (32b) with 
 $\overline{b}$, $\overline{c}$, $\overline{c}_k$ given by eq. (4) once $x_k$ is computed in terms of $T_n$ from eq. (44). As is apparent upon comparing eqs. (19,45), the solution for $x_k$ is $x_k = \frac{T_{k-1}}{T_0}$.
 
If there are two couplings $g_1$, $g_2$ then the general form of $R$ is 
\begin{equation}\tag{46}
R = \sum_{m=1}^\infty \sum_{n=0}^m \sum_{k=0}^{m-1} T_{m,n;k} g_1^{m-n} g_2^n L^k
\end{equation}
where $L = \ln \left( \frac{\mu}{Q}\right)$ and $g_1, g_1$ satisfy eq. (22) so that $g_a = g_a\left( ln \frac{\mu}{\Lambda}\right)$.  Since $R$ is independent of the unphysical renormalization mass scale $\mu$, then 
\begin{equation}\tag{47}
\mu \frac{d}{d\mu} R = \left( \mu \frac{\partial}{\partial \mu} + \beta_a \frac{\partial}{\partial g_a}\right) \sum_{k=0}^\infty A_k (g_1,g_2) L^k =0
\end{equation}
where
\begin{equation}\tag{48}
A_k = \sum_{m=0}^\infty \sum_{n=0}^{m+k+1} T_{m+k+1,n;k} g_1^{m+k+1-n} g_2^n.
\end{equation}
By eqs. (47) and (22), we find that
\begin{equation}\tag{49}
A_{k+1}\left(g_a\left( \ln\frac{\mu}{\Lambda}\right)\right) = \frac{-1}{k+1} \frac{d}{d\left( \ln \frac{\mu}{\Lambda}\right)} A_k \left(g_a\left(\ln \frac{\mu}{\Lambda}\right)\right)
\end{equation}
so that $R$ in eq. (46) becomes
\begin{equation}\tag{50}
R = \sum_{k=0}^\infty \frac{(-L)^k}{k!} \left(\frac{d}{d\left( \ln \frac{\mu}{\Lambda}\right)}\right)^k  A_0 \left(g_a\left( \frac{\mu}{\Lambda}\right)\right) \nonumber  =  A_0 \left(g_a\left( \frac{Q}{\Lambda}\right)\right).
\end{equation}
As in eq. (18), all implicit and explicit dependence on $\mu$ has cancelled once the RG has been used to sum the logarithmic contributions to $R$.

In analogy with eq. (38) we now have
\begin{equation}\tag{51}
\frac{dR}{dx_{mn}^a} = 0 = \left( \frac{\partial}{\partial x_{mn}^a} + \overline{B}_{a;m,n}^b (\overline{g}_b)  \frac{\partial}{\partial \overline{g}^b}\right) \sum_{k=0}^\infty \sum_{\ell=0}^{k+1} 
\overline{T}_{k+1,\ell ;0} (\overline{g}_1)^{k+1-\ell} (\overline{g}_2)^\ell .
\end{equation}
Using eq. (36) for $\overline{B}_{a;m,n}^b$, eq. (51) becomes (with $\bar{T}_{m,n}\equiv\bar{T}_{m,n;0}$)
\begin{align}\tag{52}
\sum_{k=0}^\infty \sum_{\ell = 0}^{k+1} \Bigg\{ \frac{\partial \overline{T}_{k+1,\ell}}{\partial x_{mn}^a} & (\overline{g}_1)^{k+1-\ell} (\overline{g}_2)^\ell + 
\overline{T}_{k+1,\ell} \Big[ \overline{B}^1_{a; m,n} (k + 1 - \ell)
(\overline{g}_1)^{k-\ell} (\overline{g}_2)^\ell \nonumber \\
&+ \overline{B}_{a; m,n}^2 (\ell) (\overline{g}_1)^{k+1-\ell}
(\overline{g}_2)^{\ell-1} \Big]\Bigg\} = 0.\nonumber
\end{align}
From eq. (52) it follows
\begin{equation}\tag{53}
\frac{\partial \overline{T}_{1\ell}}{\partial x_{mn}^a} = 0 \quad (\ell = 0,1)
\end{equation}
\begin{equation}\tag{54a}
\frac{\partial \overline{T}_{20}}{\partial x_{mn}^a} + \overline{T}_{10} \left(\delta_1^a \delta_{m2} \delta_{n0} \right) + \overline{T}_{11} \left(\delta_2^a \delta_{m2} \delta_{n0} \right) = 0
\end{equation}
\begin{equation}\tag{54b}
\frac{\partial \overline{T}_{22}}{\partial x_{mn}^a} + \overline{T}_{11} \left(\delta_2^a \delta_{m2} \delta_{n2} \right) + \overline{T}_{10} \left(\delta_1^a \delta_{m2} \delta_{n2}\right) = 0
\end{equation}
\begin{equation}\tag{54c}
\frac{\partial \overline{T}_{21}}{\partial x_{mn}^a} + \overline{T}_{10} \left(\delta_1^a \delta_{m2} \delta_{n1}\right) + \overline{T}_{11} \left(\delta_2^a \delta_{m2} \delta_{n1}\right)= 0
\end{equation}
etc. \\
with $\overline{T}_{k\ell} = T_{k\ell}$ when $x_{mn}^a = 0$.  Eqs. (53, 54) lead to
\begin{equation}\tag{55}
\overline{T}_{1\ell} = T_{1\ell}
\end{equation}
\begin{equation}\tag{56a}
\overline{T}_{20} = T_{20} - x_{20}^1 T_{10} - x_{20}^2 T_{11}
\end{equation}
\begin{equation}\tag{56b}
\overline{T}_{22} = T_{22} - x_{22}^2 T_{11} - x_{22}^1 T_{10}
\end{equation}
\begin{equation}\tag{56c}
\overline{T}_{21} = T_{21} - x_{21}^1 T_{10} - x_{21}^2 T_{11}
\end{equation}
etc.\\
It is evident that $x_{mn}^a$ can be selected so that $\overline{T}_{mn} (m \geq 2)$
are all zero so that $R$ is given by just two terms
\begin{equation}\tag{57}
R = T_{10} \overline{g}_1\left(\ln \frac{Q}{\Lambda}\right) + T_{11} 
\overline{g}_2\left(\ln \frac{Q}{\Lambda}\right) 
\end{equation}
and no higher powers of $\overline{g}_a$ contribute to $R$.  The functions $\overline{\beta}_a(\overline{g}_b)$ that govern the evolution of $\overline{g}_a$ with $\ln \frac{Q}{\Lambda}$ can be found using eq. (27) once $x_{ij}^a$ has been determined.

\section{Discussion}

In this paper we have considered some aspects of RS ambiguities when using mass independent renormalization in a theory in which there are no physical mass scales and two coupling constants.  Unlike what happens when there is but one coupling, the $\beta_a$-functions that dictate how the couplings vary with the renormalization mass scale $\mu$ when there are two couplings are ambiguous at two loop order (and beyond).  Furthermore, these ambiguities do not permit one to vary the coefficients of the expansions of these functions independently, making these coefficients unsuitable for characterizing a RS when using mass-independent renormalization.  Instead, it is convenient to parameterize a RS by directly using the coefficients of an expansion of the couplings used in a ``new'' RS in terms of the couplings used in a base RS.  

A change in RS can affect the perturbative expansion for a physical quantity R in powers of the coupling. When there is a single coupling a, one can change the coefficients $c_i$ $(i\geq 2)$ in eq. (1) by a renormalization of the form of eq. (2), as is apparent from eq. (4). This means that one can characterize a RS by the values of $c_i$ $(i\geq 2)$. If one chooses a RS in which $c_i=0$ $(i\geq 2)$ then the power series for $\beta (a)$ in eq. (1) terminates (the 't Hooft scheme [13,14]) and the behavior of the running coupling found exactly in terms of a Lambert function. A second choice of $c_i$ $(i\geq 2)$ can be made using eq. (21) so that only $T_0$ and $T_1$ in the expansion of eq. (19) is non-zero, which means that the perturbative expansion for R in powers of a terminates.

A different situation arises when there are two couplings, $g_1$ and $g_2$. In this case, the expansion coefficients $c^a_{ij}$ in eq. (22) cannot be used to characterize a RS as a renormalization like that of eq. (24) does not allow all of the $c^a_{ij}$ to independently vary, as can be seen by eq (27).

It is, however, possible to use the coefficients $x_i$ of eq. (2) (when there is one coupling) and the coefficients $x^a_{ij}$ of eq. (24) (when there are two couplings) to characterize how a change of RS from some "base scheme" (such as minimal subtraction) can be affected. In the former case, a choice of the $x_i$ so that $c_i=0$ $(i\geq 2)$ can be made, while in the latter case it is not in general possible to choose the $x^a_{ij}$ so that the expansion of eq. (22) is finite. However, in both the cases of one and two couplings, the $x_i$ and $x^a_{ij}$ respectively can be chosen so that the perturbative expansion for a physical quantity R in powers of the coupling terminates, as can be seen from eqs. (45,47). With such a choice of renormalization, the expansion coefficients of the $\beta$ function ($c_i$ and $c^a_{ij}$) are now dependent on the physical quantity being considered and all higher order loop effects are absorbed into the behavior of the running coupling.

The fixed point in such a RS is clearly important. In ref. [16] the behavior of the running couping $a$ when the quantity R in eq. (19) is the total cross section ($e^+e^-\rightarrow$hardrons) is discussed. There it is shown that in a RS in which $T_n=0$ ($n\geq 2$), the four-loop contribution to $\beta (a)$ results in an infrared fixed point and a well defined low energy limit for R. Since the perturbative series for R terminates, its convergence need not be considered. It would be quite interesting to see if fixed points arise in models with more than one coupling when a finite series is used to compute particular physical quantities.

It would be appropriate to extend these considerations to the case in which there are more than two couplings (for example, the anomalous magnetic moment of the muon in the Standard Model [25]) using the techniques outlined in the preceeding section. The RS ambiguities associated with a mass parameter in models with more than one coupling is also of interest. A study of how the presence of several renormalization scales [17] affects this discussion would also be worthwhile\vspace{.5cm}.

\section*{Acknowledgements}
\noindent
F.A. Chishtie was most helpful in formulating these ideas in this paper.  Roger Macleod had several thoughtful suggestions.

\section*{Appendix - Evolution of the running couplings}

When there is one coupling $a(\mu)$ whose evolution under changes in the renormalization mass scale $\mu$ is given by eq. (1), the relationship between $a(\mu)$ and $a(\overline{\mu})$ can be found using eqs. (9-11).  In this appendix we consider the same problem when there are two couplings $g_a(\mu) (a = 1,2)$ that satisfy eq. (24).  We begin by making the expansion
\begin{equation}\tag{A.1}
g_a(\overline{\mu}) = g_a(\mu) +\sum_{i=2}^\infty \sum_{j=0}^i \sum_{k=1}^{i-1} 
\sigma_{i,j;k}^a  g_1^{i-j} (\mu) g_2^j (\mu) \ell^k.\quad
(l \equiv ln(\frac{\mu}{\bar{\mu}}))
\end{equation}
It follows from the condition
\begin{align}\tag{A.2}
\mu \frac{dg^a(\overline{\mu})}{d\mu} &= 0\nonumber \\
& = \left( \mu \frac{\partial}{\partial \mu} +  \beta_b (g_1,g_2) \frac{\partial}{\partial g_b}\right)g^a(\overline{\mu})\nonumber
\end{align}
where $\beta_b$ is given by eq. (24).  Substitution of eq. (A.1) into eq. (A.2) results in
\begin{equation}\tag{A.3a-c}
\sigma^1_{20,1} = -c^1_{20}, \qquad
\sigma^1_{21,1} = -c^1_{21},  \qquad \sigma^1_{22,1} = -c^1_{22}
\end{equation}
\begin{equation}\tag{A.4a-d}
\sigma^1_{30,1} = -c^1_{30}, \qquad
\sigma^1_{31,1} = -c^1_{31},  \qquad \sigma^1_{32,1} = -c^1_{32}, \qquad\sigma^1_{33,1} = -c^1_{33} 
\end{equation}
\begin{align}\tag{A.5a-d}
\sigma^1_{30,2} &= \frac{1}{2}\left[2(c^1_{20})^2 + c_{21}^1c_{20}^2\right], \quad
\sigma^1_{31,2} = \frac{1}{2}\left[3c^1_{21}c_{20}^1  + c_{21}^1c_{21}^2 + 2c_{22}^1 c_{20}^2\right]\nonumber \\
& \sigma^1_{32,2} = \frac{1}{2}\left[(c^1_{21})^2 + 2c_{22}^1c_{20}^1 + c_{21}^1c_{22}^2 + 2c_{22}^1 c_{11}^2\right]\nonumber \\
& \sigma^1_{33,2} = \frac{1}{2}\left[c^1_{22} c_{21}^1 + 2c_{22}^1c_{22}^2 \right].\nonumber
\end{align}
The values of $\sigma_{4j,k}^a$ can similarly be computed in terms of $c_{ij,k}^a$. We note that since eqs. (21, A.1) are symmetric between $g_1$ and $g_2$, we have symmetry in  $(c_{ij}^1, c_{i,i-j}^2)$ and $(\sigma_{i,j;k}^1, \sigma^2_{i,i-j;k})$.  Computing all of the coefficients $\sigma_{i,j;k}^a$ amounts to integrating eq. (24) with a fixed boundary value for $g^a(\mu)$.


\begin{thebibliography}{99}
\bibitem{1} D.G.C. McKeon and T.N. Sherry, \textit{Phys. Rev.} \textbf{D35}, 3854 (1987).
\bibitem{2} E.C.G. Stueckelberg and A. Peterman, \textit{Helv. Phys. Acta} \textbf{26}, 499 (1953).
\bibitem{3} M. Gell-Mann and F.E. Low,  \textit{Phys. Rev.} \textbf{95}, 1300 (1954).
\bibitem{4} N.N. Bogoliubov and D.V. Shirkov, \textit{Nuovo Cimento} \textbf{3}, 845 (1956).
\bibitem{5} G. 't Hooft, \textit{Nucl. Phys.} \textbf{B61}, 455 (1973).
\bibitem{6} S. Weinberg,  \textit{Phys. Rev.} \textbf{D8}, 3497 (1973).
\bibitem{7} J. Collins and A. MacFarlane, \textit{Phys. Rev.} \textbf{D10}, 1201 (1974).
\bibitem{8} P.M. Stevenson, \textit{Phys. Rev.} \textbf{D23}, 2916 (1981).
\bibitem{9} D.G.C. McKeon, \textit{Phys. Rev.} \textbf{D92}, 045031 (2015).
\bibitem{10} F.A. Chishtie, D.G.C. McKeon and T.N. Sherry, \textit{Phys. Rev.} \textbf{D94}, 054031 (2016).
\bibitem{11} F.A. Chishtie and D.G.C. McKeon, hep-ph 1610.06487 (\textit{Can. J. Phys.} in press).
\bibitem{12} J. Collins, ``Renormalization'' (Cambridge U. Press, Cambridge 1984).
\bibitem{13} G. `t Hooft, ``The Whys of Subnuclear Physics'' Erice 1977, Edited by A. Zichichi (Plenum, New York, 1979).
\bibitem{14} T. Banks and A. Zaks,  \textit{Nucl. Phys.} \textbf{B196}, 189 (1982).
\bibitem{15} C. Ford, D.R.T. Jones, P.W. Stephenson and M.B. Einhorn,  \textit{Nucl. Phys.} \textbf{B395}, 17 (1993)
\bibitem{16} F.A. Chishtie and D.G.C. McKeon, \textit{Phys. Rev.} \textbf{D95}, 116013 (2017).
\bibitem{17} T. Steele, Zhi-Wei Wang, and D.G.C. McKeon, \textit{Phys. Rev.} \textbf{D90}, 105012 (2014).
\bibitem{18} I.M. Gessel, \textit{Jour. of Comb. Th.} A, \textbf{45}, 178 (1987).
\bibitem{19} J.F. Fortin, B. Grinstein and A. Stergiou, \textit{JHEP} \textbf{07}, 025 (2012).
\bibitem{20} S. Borowka, T. Hahn, S. Heinemeyer, G. Heinrich and W. Hollik, \textit{Eur. Phys. J.} \textbf{C74}, 2994 (2014).
\bibitem{21} G. Degrassi, S. Di Vita and P. Slavich, \textit{Eur. Phys. J.} \textbf{75}, 61 (2015).
\bibitem{22} S. Borowka, T. Hahn, S. Heinemeyer, G. Heinrich and W. Hollik, \textit{Eur. Phys. J.} \textbf{C75}, 424 (2015).
\bibitem{23} T.A. Ryttov and R. Shrock, \textit{Phys. Rev.} \textbf{D86}, 065032 (2012); ibid. \textit{Phys. Rev.} \textbf{D86}, 085005 (2012).
\bibitem{24} R. Shrock, \textit{Phys. Rev.} \textbf{D88}, 036003 (2013); ibid. \textit{Phys. Rev.} \textbf{D90}, 045011 (2014).
\bibitem{25} Tatsumi Aoyama, Masashi Hayakawa, Toichiro Kinoshita, and Makiko Nio, \textit{Phys. Rev. Lett.} \textbf{109}, 111808 (2012); F. Jegerlehner, A. Nyffeler, \textit{Phys. Rep.} \textbf{477}, 1 (2009).
\bibitem{26} W. Celmaster and R.J.Gonsalves, \textit{Phys. Rev.} \textbf{D21}, 3112 (1980).
\bibitem{27} J. A. Gracey, \textit{Phys. Rev.} \textbf{D90}, 094026 (2014).
\bibitem{28} F. A. Chrishtie, D. G. C. McKeon and T. N. Sherry (in preparation).
\end{thebibliography}
\end{document}